\documentclass[%12pt,%
twocolumn,
 amsmath,amssymb,
 aps,
pra,
 showpacs
]{revtex4-1}
\usepackage{placeins}
\usepackage{amsmath}
\usepackage{mathrsfs}
\usepackage{graphicx}% Include figure files
\usepackage{dcolumn}% Align table columns on decimal point
\usepackage{bm}% bold math
\usepackage{ulem}
 \usepackage{relsize}
\usepackage[%Uncomment any one of the following lines to test 
margin=.7in,
]{geometry}

\begin{document}
%\affiliation
\title{Complicated high-order harmonic generation due to the falling edge of a trapezoidal laser pulse
}

\author{H. Ahmadi$^{1}$}
\author{M. Vafaee$^{2}$}
\author{A. Maghari$^{1}$}
%\email {Corresponding author: m.vafaee@modares.ac.ir}

\affiliation{
$^{1}$Department of Physical Chemistry, School of Chemistry, College of Science, University of Tehran, Tehran, Iran
\\$^{2}$Department of Chemistry, Tarbiat Modares University, P. O. Box 14115-175, Tehran, Iran }

\begin{abstract}
High-order harmonic generation (HHG) is investigated for H$_2^+$ and its isotopomers under seven- and ten-cycle trapezoidal laser pulses at 800 nm wavelength and $I$=4$\times 10^{14}$ W$/$cm$^2$ intensity. We solved numerically full-dimensional electronic time-dependent Schr\"{o}dinger equation with and without the Born-Oppenheimer approximation. We show that the HHG at the falling edge of a trapezoidal laser pulse can result in redshift and complexity on the total HHG spectrum which can be removed by considering  different laser pulse duration and nuclear motion not possible for sin$^2$ and Gaussian laser pulses.  We resolve the redshifts and complexities of the HHG spectra into different electronic and vibrational states and their interferences. 
\end{abstract}

%\vspace{2pc}
%\small{\textbf{Keywords:}  Time-dependent Schr\"{o}dinger equation, High-order harmonic generation, Beyond dipole approximation, Super-intense xuv ultrashort laser, Beyond Born-Oppenheimer approximation.}
%\submitto{\JPB}
%\twocolumn
%\title{}
\maketitle
%\twocolumn
\section{Introduction}
 The non-perturbative interaction of intense laser pulses with atoms and molecules gives rise to the generation of high energetic photons well known as high-order harmonic generation (HHG) [1,2]. The HHG is well understood by a three-step model proposed by Corkum [3] and extended by Lewenstein \textit{et al}. [4]. First, an electron tunnels into continuum from suppressed potential created with combinations of  system's Coulomb potential and laser pulse field. Then, the electron oscillates in the laser field and moves away from the ion core, and after the field reverses, it comes back to the core. Finally, the electron may recombine with its parent ion, leading to photon emission of  high frequencies which are multiples of the frequency of the driving laser.  The HHG is used to produce single isolated or trains of attosecond laser pulses which are necessary for the real-time observation of electronic dynamics [5].
The emitted photons in the HHG possess also time-dependent structural information on sub-femtosecond time scales which can be used to obtain the details of a system under investigation [6-10]. For example, from  amplitude [11] and frequency [12] modulations in the HHG spectrum, molecular internuclear distance can be obtained.

The HHG process in molecules is more complex than that in atoms due to nuclear motion, two-center interference and different orientations of a molecule with respect to the laser field.
We mention some recent works on the effect of nuclear motion on the HHG as follows.
Experiments on H$_2$ and D$_2$ [7,13-14], and CH$_4$ and CD$_4$ [7]  show that the HHG is suppressed  more for the lighter isotopes due to faster nuclear motion. The nuclear motion can  shorten the length of attosecond laser pulse trains [15] and can produce an isolated attosecond laser pulse because of long-trajectory suppression [16,17]. It is reported that even-order harmonics are produced when superposition of two electronic states is formed for large internuclear distances as a result of the system's symmetry breaking [18]. It is also shown that in the presence of nuclear motion, like in H$_2^+$ and H$_2$, the energy difference between the two lowest bonding and anti-bonding electronic states decreases for large internuclear distances, thus anti-bonding electronic state is populated and this electronic state also contributes to the HHG [19]. The two-center interference  minimum disappears at large internuclear distances where recombination to the first excited (anti-bonding) state becomes important [20]. The suppression of long trajectories and the HHG yield at different intensities due to nuclear motion are also reported [21]. 

By considering nuclear movement and laser pulse duration, Bian \textit{et al}. observed only the redshift of HHG spectrum for a sin$^2$ laser pulse [12].
For laser pulses with sin$^2$ and Gaussian envelopes having only rising and falling parts, the blueshift or redshift of harmonics relative to odd harmonic orders  occur inevitably at the rising and falling parts, respectively (for more details, see ref. [12] and references therein). In practice, it is more difficult to observe only the redshift than the blueshift.
 At some conditions for sin$^2$ and Gaussian laser pulses, depending on pulse duration, type of the isotope and laser pulse intensity, both blueshift and redshift of harmonics can be present, leading to broadening of the harmonics if the HHG happens equally at both rising and falling edges of the laser pulse [12]. Therefore, the HHG spectrum certainly changes in the case of sin$^2$ and Gaussian laser pulses. 

In this work, we investigate the effect of nuclear movement on the HHG spectrum under trapezoidal laser pulses, mainly focused on the laser falling part. We address some  questions for different isotopes under  trapezoidal laser pulses not already introduced and answered: Would the rising and falling parts of a trapezoidal laser pulse affect the HHG? To what extent the effects would be dependent on the type of the isotope? What molecular electronic and vibrational components contribute into the effects? Finally, how could we control the shape of HHG spectrum due to the corresponding effects?

For the trapezoidal laser pulses, we show that the redshift of harmonics may occur depending on the isotope and pulse duration. It is shown that the effect of laser falling part on the HHG spectrum could be even removed and therefore controlled for a trapezoidal laser pulse which is not possible for sin$^2$ and Gaussian laser pulses as mentioned above. 

We resolve the redshifts and complexities of the HHG spectra into different electronic and vibrational states and their interferences. Similar decomposing the HHG into different electronic [19,20] and vibrational [18] states has been reported but here we apply them to resolve the observed redshifts. We show that not only individual electronic and vibrational states are involved in the redshifts but also their interferences are significant and should be taken into account. To our knowledge, resolving the redshifts into their components have  not been reported, even for sin$^2$ and Gaussian laser pulses.
To see whether the nuclear motion affect the HHG, we performed the calculations for both fixed and freed nuclei. For the latter, we used different isotopomers in order to have a better understanding  of the effects of the nuclear movement on the observed complexities in the HHG.

 Here, full-dimensional  electronic time-dependent Schr\"{o}dinger equation (TDSE) beyond the Born-Oppenheimer approximation (NBO) is solved numerically for H$_2^+$, D$_2^+$ and X$_2^+$ (X is a virtual isotope of H being 10 times heavier). The full-dimensional electronic TDSE within the Born-Oppenheimer approximation (BO) for H$_2^+$, which is indicated throughout the article by H$_2^+$(BO), is also solved to compare with NBO results. The equilibrium internuclear distance within BO is set to $R_{eq}=1.96$ a.u. Calculations have been done with seven- and ten-cycle laser pulses of 800 nm wavelength with $I=$4 $\times 10^{14}$ W$/$cm$^2$ intensity. We suppose that molecular ions are aligned in such a way that their internuclear-distance axis is parallel to the laser polarization direction. Nowadays the molecular alignment in strong laser fields is readily feasible and it is applied in several experiments [6,9,22-27]. We use atomic units throughout the article unless stated otherwise.

% $\vspace{-1cm}$
\section{Computational Methods}
The time-dependent Schr\"{o}dinger equation for H$_2^+$  (D$_2^+$) with electron cylindrical coordinate $(z,\rho)$ ͒ with respect to the molecular center of mass and internuclear distance $R$, for both $z$ and $R$ parallel to the laser polarization direction, can be read (after elimination of the center-of-mass motion and ignoring molecular rotation) as [28-29]

\begin{eqnarray}\label{eq:1}
  i \frac{\partial \psi(z,\rho, R;t)}{\partial t}={\widehat H}(z,\rho, R;t)\psi(z,\rho, R;t).
\end{eqnarray}
In this equation, \^{H} is the total electronic and nuclear Hamiltonian which is given by
\begin{eqnarray}\label{eq:2}
 \widehat{H}(z,\rho, R;t)=&\mathlarger{-\frac{2m_N+m_e}{4m_Nm_e}[\frac{\partial^2}{\partial \rho^2}+\frac{1}{\rho}\frac{\partial}{\partial \rho}+\frac{\partial^2}{\partial z^2}]}
\nonumber \\
& \mathlarger{-\frac{1}{m_N}\frac{\partial^2}{\partial R^2}+V_C(z,\rho, R;t)},
\end{eqnarray}
with
\begin{eqnarray}\label{eq:3}
 \widehat{V}_C&(z,\rho, R,t)=\mathlarger{-\frac{1}{\sqrt{(z+\frac{R}{2})^2+\rho^2}}-\frac{1}{\sqrt{(z-\frac{R}{2})^2+\rho^2}}}
\nonumber \\
 &\mathlarger{+\frac{1}{R}+(\frac{2m_N+2m_e}{2m_N+me})zE_0f(t)sin(\omega t)}.
\end{eqnarray}
In these equations, $E_0$ is the laser peak amplitude, $m_e$ and $m_N$ are  masses of the electron and single nuclei, $\omega$ angular frequency  and \textit{f}(t) is the laser pulse envelope, which rises linearly during the first two cycles, then is constant for a few cycles and decreases during the last two cycles. For example, the shape of the ten-cycle laser pulse used in this work is shown in Fig. 1.

%===========================Figure===========================================
\begin{figure}[ht]
\begin{center}
\begin{tabular}{c}
\centering
\resizebox{80mm}{50mm}{\includegraphics{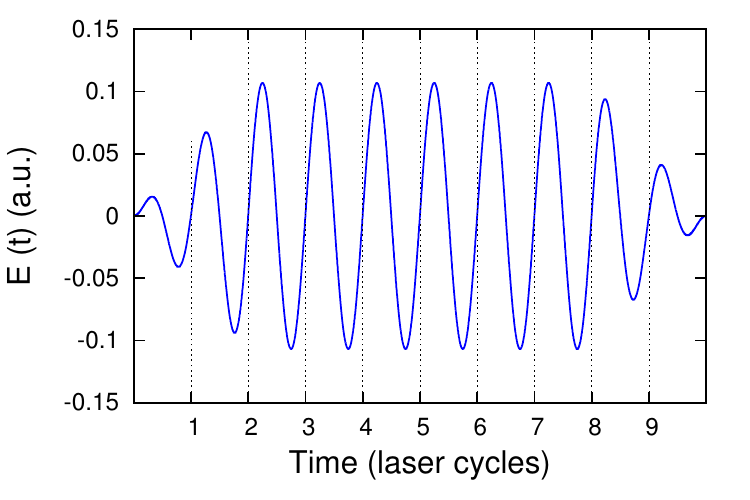}}
\end{tabular}
\caption{
\label{HHG} 
(Color online) The ten-cycle laser pulse shape with a  two-cycle linear turn-on, followed by a constant intensity, and a two-cycle linear turn-off at 800 nm wavelength ($\omega=0.057$ a.u.) and $I$=4$\times 10^{14}$ W/cm$^2$ intensity.		}
\end{center}
\end{figure}
%====================================
%===========================Figure===========================================
\begin{figure*}[ht]
%\begin{}
%\begin{tabular}{c}
\centering
%\captionsetup{justification=centering}
\resizebox{175mm}{55mm}{\includegraphics{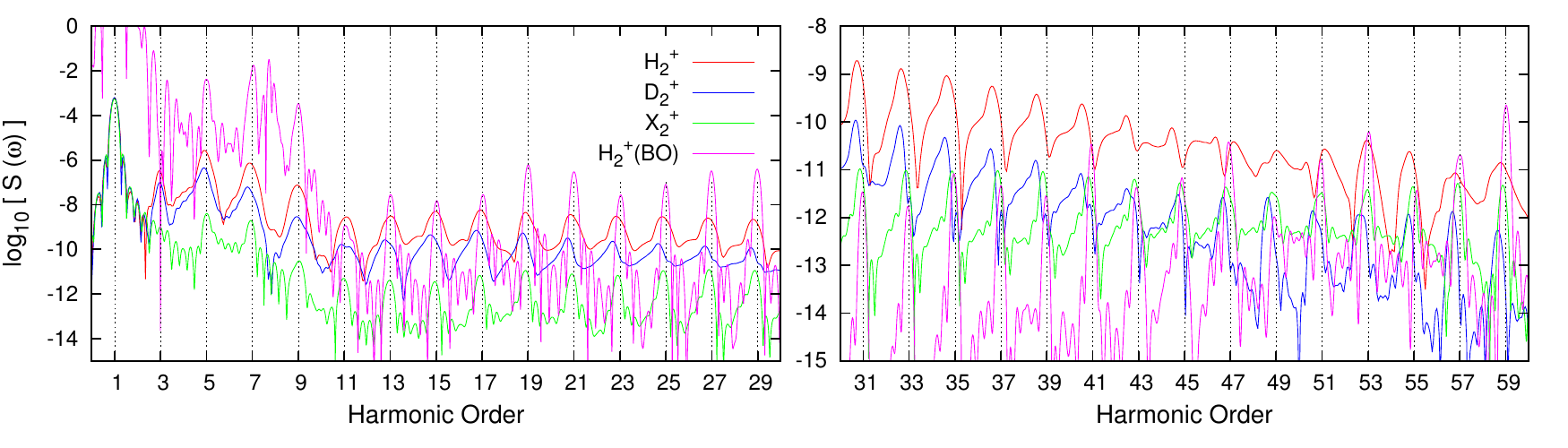}}
%\end{tabular}
\caption{
%\label{PIR}
(Color online) High-order harmonic spectra produced by the NBO H$_2^+$, D$_2^+$, X$_2^+$  and BO H$_2^+$ (H$_2^+$(BO)) under seven-cycle laser pulses of 800 nm wavelength at $I=$4 $\times 10^{14}$ W$/$cm$^2$ intensity. For better clarity, the range of 1-29 and 31-59 harmonics of the spectra are shown in the left and right panels, respectively.
}
%\end{center}
\end{figure*}
%============================================
%===========================Figure===========================================
\begin{figure}[ht]
\begin{center}
\begin{tabular}{l}
\resizebox{85mm}{130mm}{\includegraphics{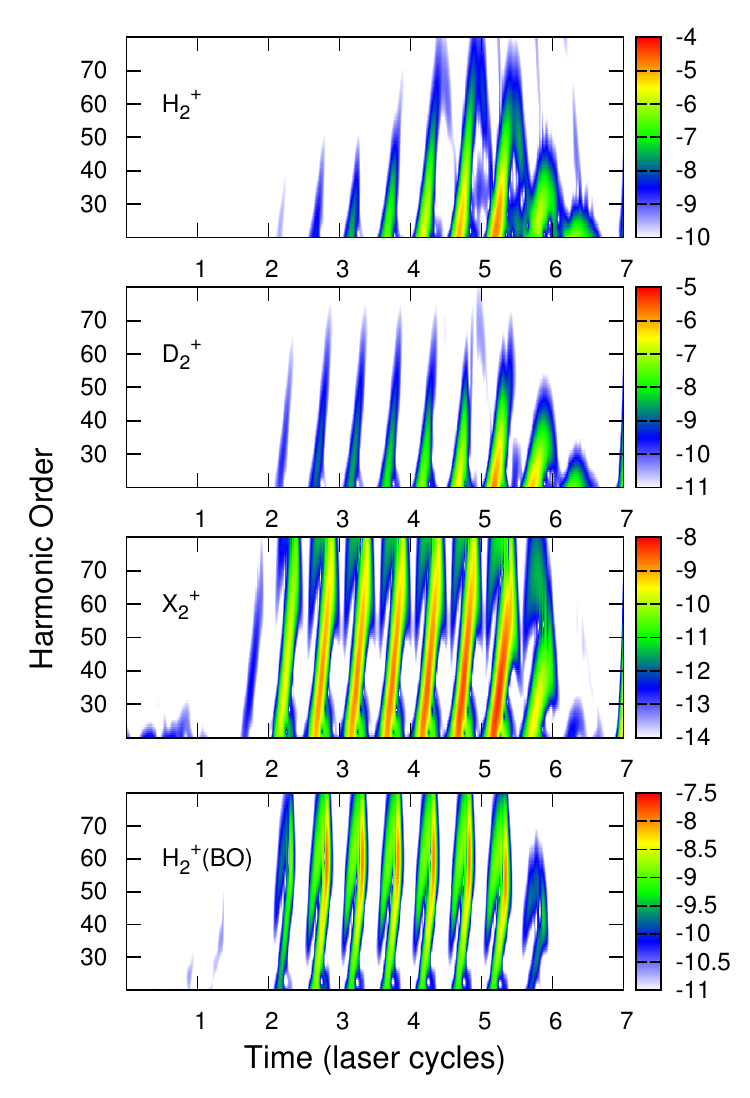}}
\end{tabular}
\caption{
\label{<z>}
(Color online) The Morlet-wavelet time profiles for NBO H$_2^+$, D$_2^+$ and X$_2^+$ and BO H$_2^+$ (H$_2^+$(BO)) under seven-cycle laser pulses of 800 nm wavelength and $I$=4 $\times 10^{14}$ W$/$cm$^2$ intensity. The HHG intensities are depicted in color logarithmic scales on the right side of panels.		
		}
\end{center}
\end{figure}
%===================================

The TDSE is solved using unitary split-operator methods [30-31] with 11-point finite difference scheme through a general nonlinear coordinate transformation for both electronic and nuclear coordinates which is described in more details in our previous works [32-34]. The grid points for z, $\rho$, and R coordinates are 500, 83, and 210, respectively. The finest grid size values in this
adaptive grid schemes are 0.13, 0.1, and 0.025, respectively for $z$, $\rho$, and $R$ coordinates. The grids extend up to $z_{max}
= 98$, $\rho_{max} = 25$, and $R_{max}= 16$. 
The HHG spectra are calculated as the square of the windowed Fourier transform of dipole acceleration $a_z(t)$ in the electric field direction (z) as
\begin{eqnarray}\label{eq:4}
  S(\omega)=\mathlarger{\vert}\int_0^T<\psi|a_z(t)\mid \psi>\,H(t)\,exp[-i\omega t]\,dt\; \mathlarger{\vert} ^2,
\end{eqnarray}
where
\begin{eqnarray}\label{eq:5}
  H(t)= \frac{1}{2}[1-cos(2\pi \frac{t}{T})],
\end{eqnarray}
is the Hanning filter and $T$ is the total pulse duration (one optical cycle of 800 nm wavelength equals 2.6 fs). 

To obtain contributions of different vibrational and electronic states to total HHG spectrum, we first decompose the total wavefunction as [20]
\begin{eqnarray}\label{eq:6}
  &\psi(z,\rho,R;t)=\sum_{i=1}^{4}
   c_i(R;t)\psi_i(z,\rho;R)+\psi_{res}(z,\rho,R;t).\nonumber \\
\end{eqnarray}
$\psi_{i}(z,\rho;R)$'s are  four lowest Born-Oppenheimer electronic wavefunctions of the system and $c_{i}(R;t)= <\psi_{i}(z,\rho;R)|\psi(z,\rho,R;t)>_{z,\rho}$ which is integrated over $z$ and $\rho$ electronic coordinates. In Eq. (6), $c_{1}(R;t)=c_{g}(R;t)$, $c_{2}(R;t)=c_{u}(R;t)$, $\psi_{1}(z,\rho;R)$=$\psi_{g}(z,\rho;R)$ and $\psi_{2}(z,\rho;R)$=$\psi_{u}(z,\rho;R)$ in which $\psi_{g}(z,\rho;R)$ and $\psi_{u}(z,\rho;R)$ are ground and first excited electronic wavefunctions, respectively, corresponding to the $1s\sigma _g$ and $2p\sigma_u$ states. 
If we substitute Eq. 6 to Eq. 4 and retain the dominant terms, we arrive at [20]
\begin{eqnarray}\label{eq:7}
  S_{tot}\simeq S_g(\omega)+S_u(\omega)+2[A_g^*(\omega)\times A_u(\omega)],
\end{eqnarray}
where $S_g(\omega)=|A_g(\omega)|^2$ and $S_u(\omega)=|A_u(\omega)|^2$ and
\begin{eqnarray}\label{eq:8}
  A_g(\omega)=&\\
  \int 2Re<c_g(R;t)&\psi_g(z,\rho;R)\mid a_z(t)\mid \psi_{res}(z,\rho,R;t)> e^{-i\omega t} dt,
   \nonumber \\
    A_u(\omega)=&\nonumber \\
    \int 2Re<c_u(R;t)&\psi_u(z,\rho;R)\mid a_z(t)\mid \psi_{res}(z,\rho,R;t)> e^{-i\omega t} dt.
    \nonumber
 \end{eqnarray}
 $S_g(\omega)$ and $S_u(\omega)$ denote recombination to the $1s\sigma_g$ and $2p\sigma_u$ states, respectively and the term $2[A_g^*(\omega)\times A_u(\omega)]$ corresponds to the electronic interference term (EIT) between these two electronic states [20]. In this work, the second ($\psi_{3}(z,\rho;R)$) and third ($\psi_{4}(z,\rho;R)$) excited electronic states are not considerably populated (both reach maximum $\sim$ 1.4\% while the population of the first excited state reaches up to 40\%) during the interaction and therefore the corresponding terms $S_3(\omega)$ and $S_4(\omega)$ are negligible and consequently have been ignored. 
 
 We also decompose the $S_g(\omega)$ into the bound and dissociative parts as
 \begin{eqnarray}\label{eq:9}
  S_g(\omega)= S_b(\omega)+S_d(\omega)+2[A_b^*(\omega)\times A_d(\omega)],
\end{eqnarray}
in which the bound term $S_b(\omega)=S_b^{n_{max}}(\omega)$ and $A_b(\omega)=A_b^{n_{max}}(\omega)$ in which $S_b^n(\omega)=|A_b^n(\omega)|^2$  with
\begin{eqnarray}\label{eq:10} 
  A_b^n(\omega)=&  \\
   \sum_{\nu_i=0}^{n-1}\int 2Re<c_{\nu_i}(t)\psi_{\nu_i}(R)\psi_g(z,\rho;R)&\mid a_z(t)\mid \psi_{res}(t)>\nonumber \\ e^{-i\omega t} dt,\nonumber
\end{eqnarray}
which is sum over the vibrational states with vibrational quantum number $\nu_i$ and $c_{\nu_i}(t)=<\psi_{\nu_i}(R)|c_{g}(R;t)>$. The dissociative term $S_d(\omega)=|A_d(\omega)|^2$ with $A_d(\omega)=A_g(\omega)-A_b(\omega)$. The last term in the Eq. (9), $2[A_b^*(\omega)\times A_d(\omega)]$, corresponds to the vibrational interference term (VIT) between the bound and dissociative terms of the $1s\sigma_g$ state.
The time profile of harmonics is obtained by Morlet-wavelet transform of dipole acceleration $a_z(t)$ via [35-36]
\begin{eqnarray}\label{eq:12}
  &\mathlarger{w(\omega,t)= \sqrt{ \frac{\omega}{\pi^\frac{1}{2}\sigma}}\times}
 \nonumber \\
 &\mathlarger{\int_{-\infty}^{+\infty}a_z(t^\prime)exp[-i\omega (t^\prime-t)]exp[-\frac{\omega^2 (t^\prime-t)^2}{2\sigma^2}]dt^\prime.}
\end{eqnarray}
We set $\sigma=2\pi$ in this work as used in [36].
  
%===========================Figure===========================================
\begin{figure*}[ht]
\begin{center}
\begin{tabular}{l}
\resizebox{175mm}{55mm}{\includegraphics{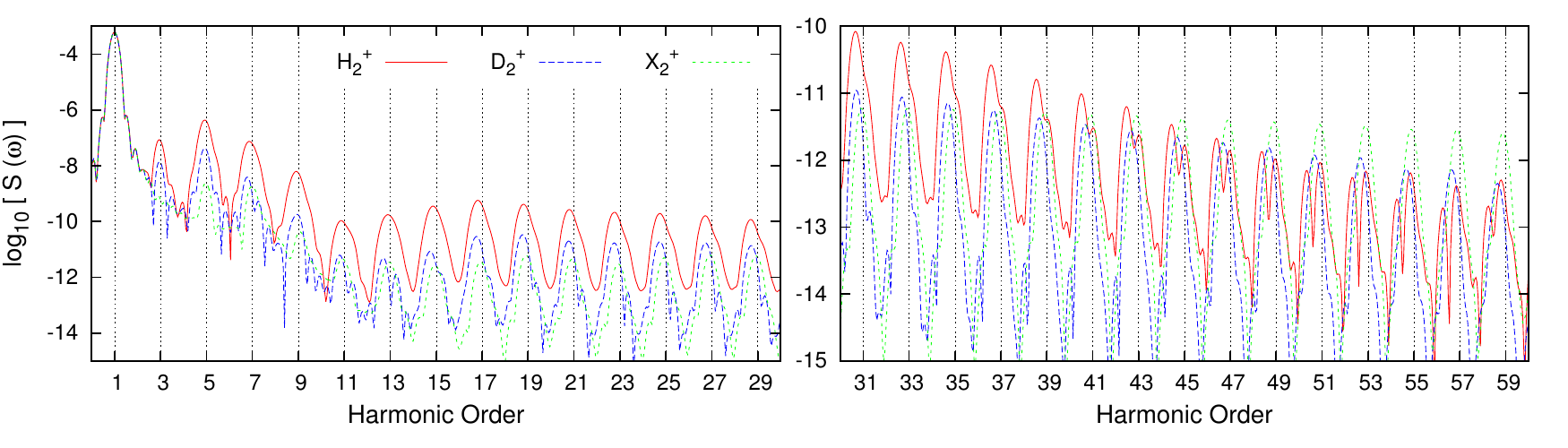}}
\end{tabular}
\caption{
\label{<z>}
(Color online) The same as FIG. 2, but Fourier transform is done over first five optical cycles (set $T=5$ o.c. in Eq. (4)).		
		}
\end{center}
\end{figure*}
%===================================

%===========================Figure===========================================
\begin{figure*}[ht]
\begin{center}
\begin{tabular}{l}
\resizebox{175mm}{55mm}{\includegraphics{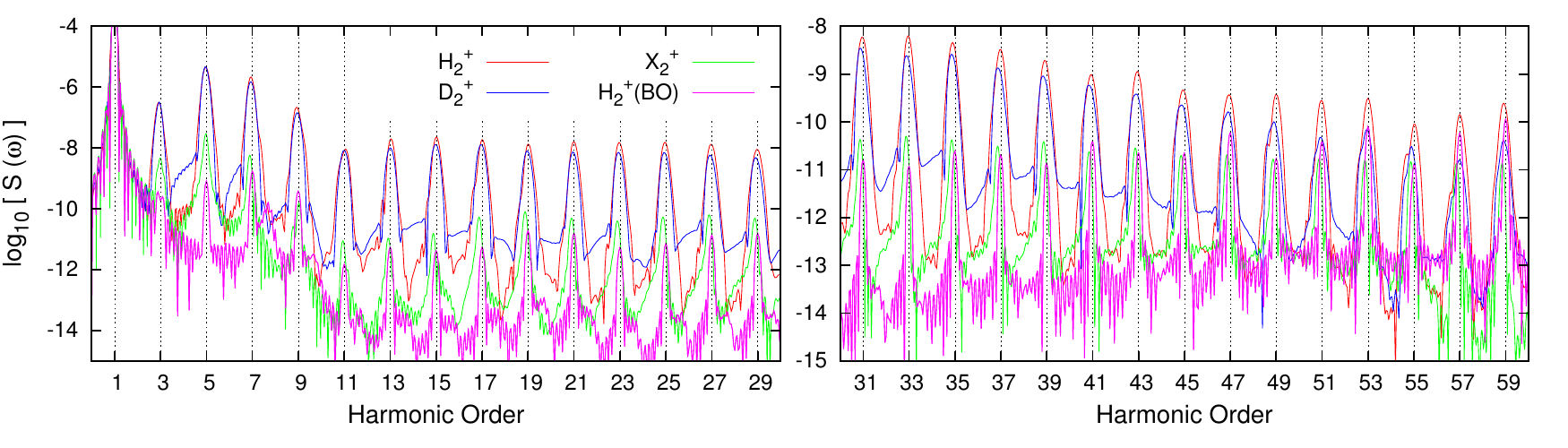}}
\end{tabular}
\caption{
\label{<z>}
(Color online) The same as FIG. 2, but for a ten-cycle laser pulse shown in Fig. 1.		
		}
\end{center}
\end{figure*}
%===================================

%===========================Figure===========================================
\begin{figure}[ht]
\begin{center}
\begin{tabular}{l}
\resizebox{85mm}{130mm}{\includegraphics{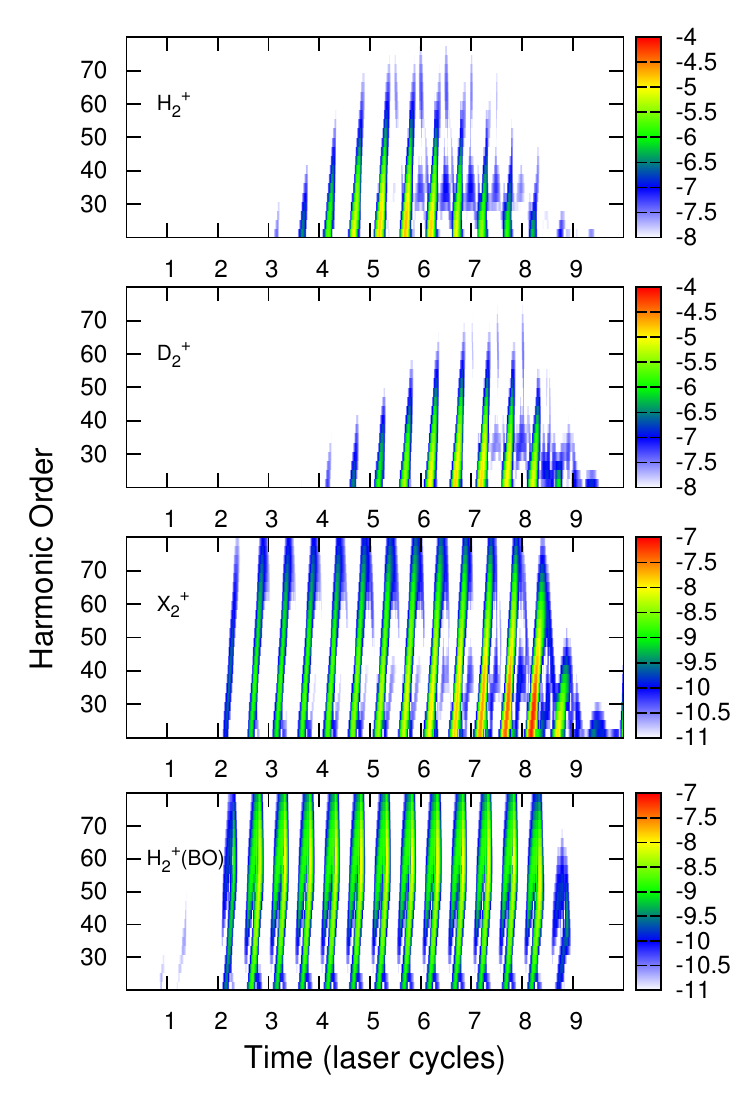}}
\end{tabular}
\caption{
\label{<z>}
(Color online) The same as FIG. 3, but for a ten-cycle laser pulse.		
		}
\end{center}
\end{figure}
%===================================
%===========================Figure===========================================
\begin{figure}[t]
\begin{center}
\begin{tabular}{l}
\resizebox{80mm}{50mm}{\includegraphics{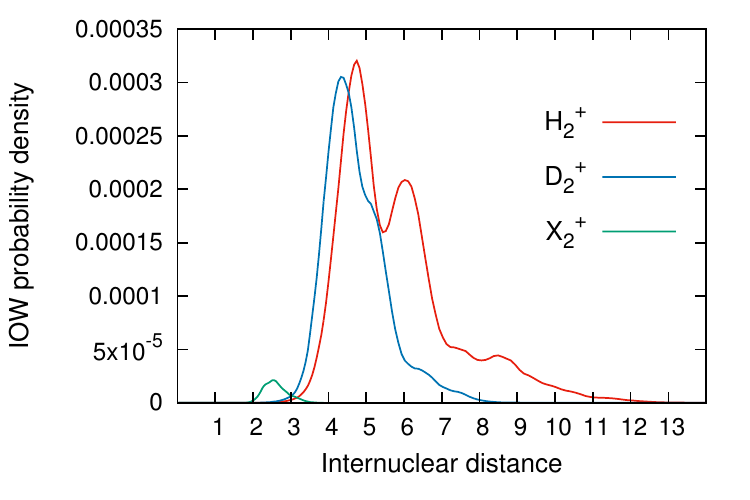}}
\end{tabular}
\caption{
\label{<z>}
(Color online) The ionized outgoing wavepacket (IOW) probability density in the $z$ direction as a function of internuclear distance for NBO H$_2^+$, D$_2^+$ and X$_2^+$ under ten-cycle laser pulses of 800 nm wavelength and $I$=4 $\times 10^{14}$ W$/$cm$^2$ intensity.		
		}
\end{center}
\end{figure}
%===================================
%===========================Figure===========================================
\begin{figure}[ht]
\begin{center}
\begin{tabular}{l}
\resizebox{85mm}{120mm}{\includegraphics{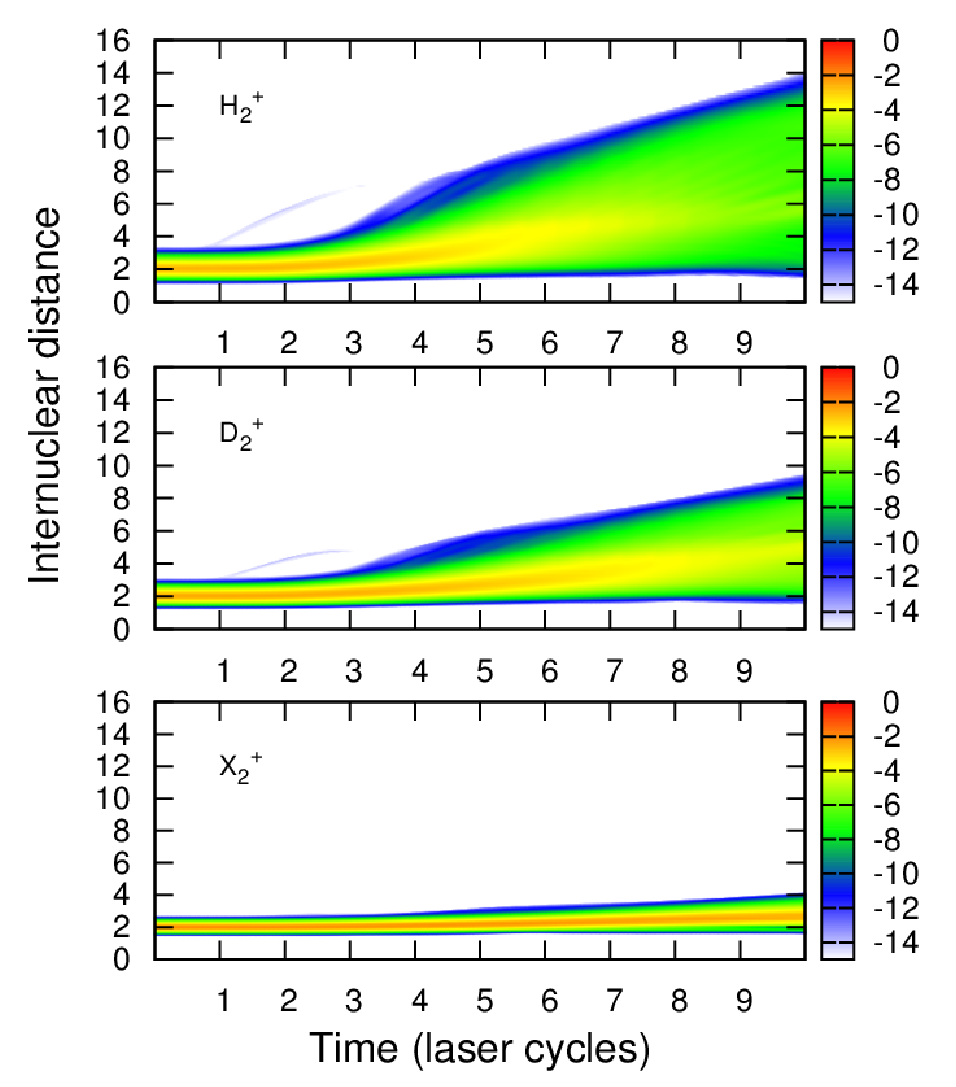}}
\end{tabular}
\caption{
\label{<z>}
(Color online) Time-dependent population for NBO H$_2^+$, D$_2^+$ and X$_2^+$ as function of internuclear distance under ten-cycle laser pulses of 800 nm wavelength and $I$=4 $\times 10^{14}$ W$/$cm$^2$ intensity.		
		}
\end{center}
\end{figure}
%===================================

%===========================Figure===========================================
\begin{figure}[ht]
\begin{center}
\begin{tabular}{c}
\centering
\resizebox{85mm}{130mm}{\includegraphics{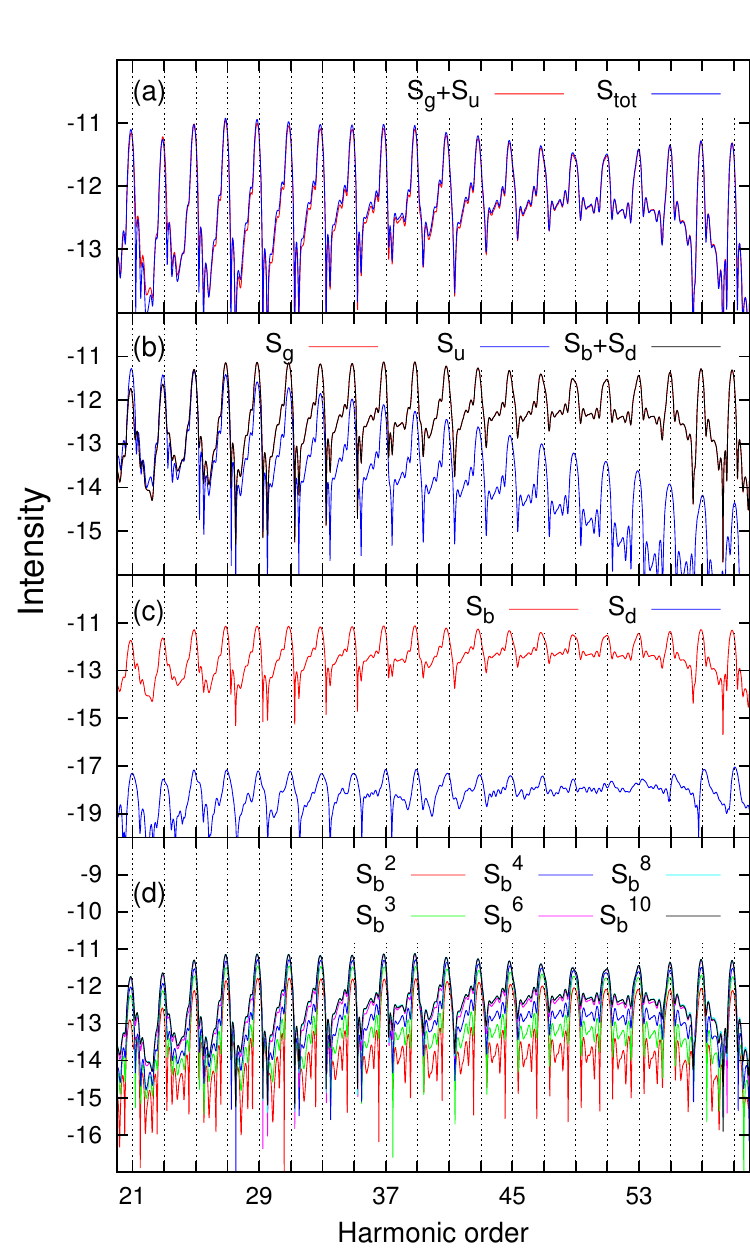}}
\end{tabular}
\caption{
\label{internuclear_distance}
(Color online) The components of the HHG spectrum of the X$_2^+$ exposed to seven-cycle laser pulses of 800 nm wavelength and $I$=4 $\times 10^{14}$ W$/$cm$^2$ intensity.
		}
\end{center}
\end{figure}
%==========================================
%===========================Figure===========================================
\begin{figure}[ht]
%\begin{center}
\centering
\begin{tabular}{c}
\centering
\resizebox{85mm}{130mm}{\includegraphics{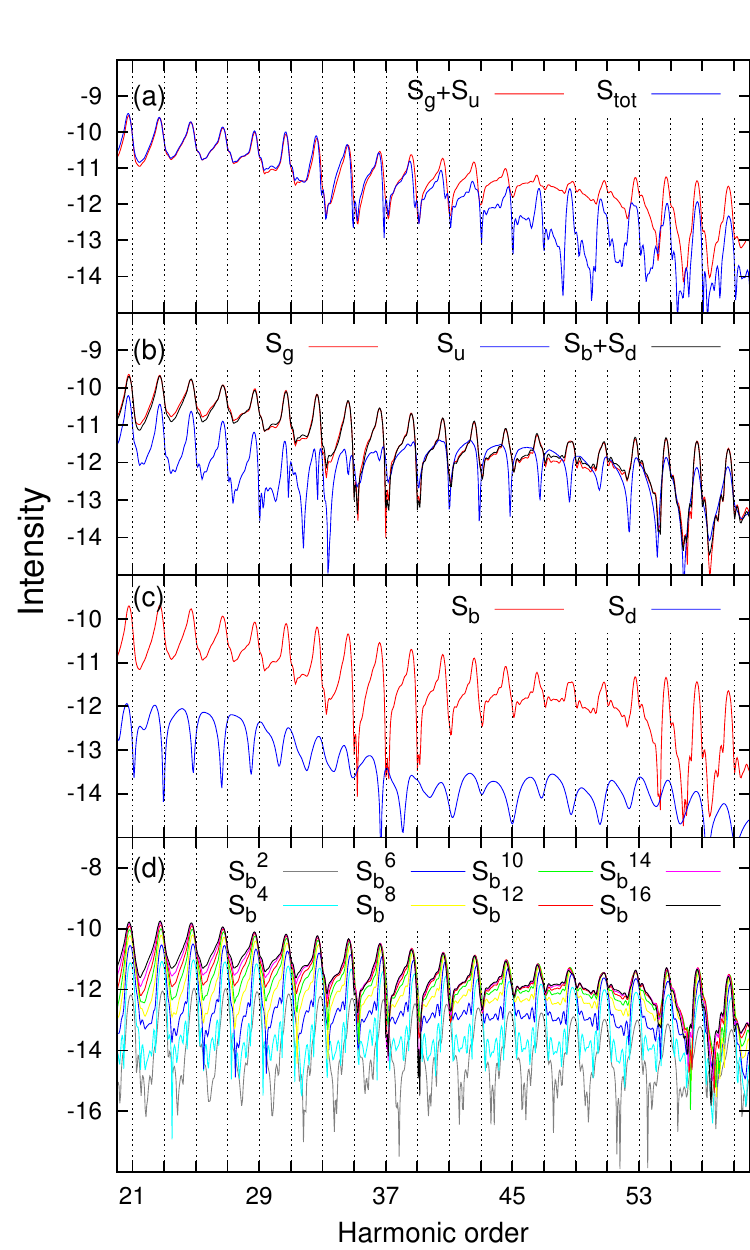}}
\end{tabular}
\caption{
\label{Fig-2}
(Color online) The same as FIG. 9, but for D$_2^+$.		
		}
%\end{center}
\end{figure}
%=================================
%===========================Figure===========================================
\begin{figure}[tt]
%\begin{center}
\centering
%\begin{tabular}{r}
\resizebox{85mm}{130mm}{\includegraphics{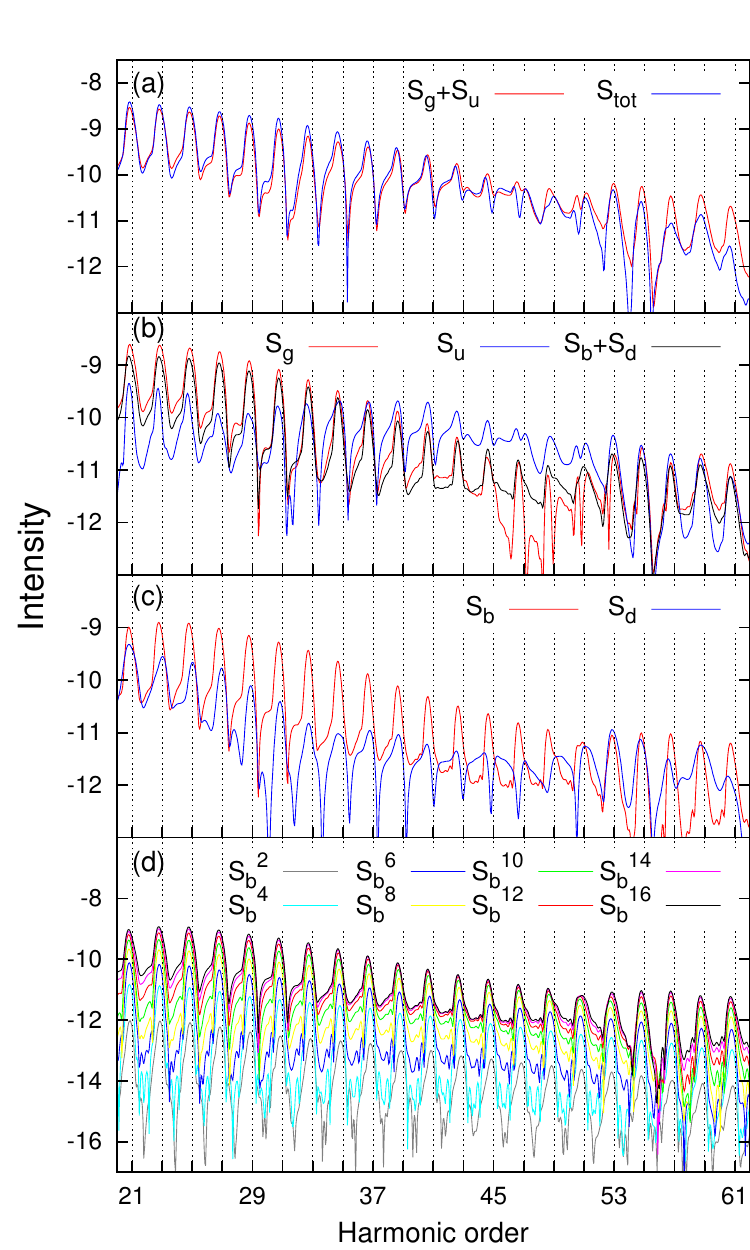}}
%\end{tabular}
\caption{
\label{<z>}
(Color online) The same as FIG. 9, but for H$_2^+$.		
		}
%\end{center}
\end{figure}
%===================================

\section{Results and Discussion}
The HHG spectra of H$_2^+$, D$_2^+$, X$_2^+$ and H$_2^+$(BO) obtained under seven-cycle 800 nm laser pulses of I=4$\times 10^{14}$  W$/$cm$^2$ intensity are shown in Fig. 2. For better visualization, harmonic orders between 1-29 and 31-59 are shown separately. For the fixed-nuclei approximation case (H$_2^+$(BO)), only odd harmonics are seen while for NBO cases, although odd harmonics are dominant at low orders but the HHG signals are complicated. For X$_2^+$ isotopomer, small redshifts in harmonics relative to odd harmonics  are seen which are more discernible for higher harmonics. For D$_2^+$ and H$_2^+$ cases, the patterns of the redshift are complicated and the HHG spectra broaden for most harmonic orders. 

To demonstrate the origin of redshifts discussed above, we plotted temporal dependence of harmonic orders for both BO and NBO cases in Fig. 3. As can be seen, mostly short trajectories contribute to the HHG because long-trajectory suppression is considerable due to the wavepacket spreading and also nuclear motion [21]. For  laser pulses of sin$^2$ and Gaussian envelopes, we can attribute rising and falling parts to the laser pulse. The blueshift and redshift of harmonics occur in rising and falling parts, respectively (see Ref. [12], and references therein). Bian \textit{et al.} showed that the HHG occurs mostly at the laser falling  part because of the nuclear motion and enhanced ionization, giving rise to the redshift of harmonics [12]. In this work, we used trapezoidal envelopes having a two-cycle rising  and a two-cycle falling part. As it is seen in Fig. 3, for H$_2^+$, D$_2^+$ and X$_2^+$, there are three relatively intense peaks between 5 to 7 laser cycles, i.e. in the falling part of the laser pulse, while there is not any significant peak between 0-2 optical cycles for the laser rising part. For X$_2^+$, we see several intense peaks before five optical cycles and therefore the redshifts from laser falling part become less important in comparison with H$_2^+$ and D$_2^+$ and for D$_2^+$ similarly the redshifts are less than those in H$_2^+$. To see whether the redshifts are due to laser falling part, we plotted the HHG spectra of three isotopomers in Fig. 4 which is similar to Fig. 2 but here Fourier transform is integrated over the first 5 optical cycles to remove the contribution of the last two-cycle falling part of the laser pulse. As it is obvious, the HHG spectra have been smoothed and complicated features, already seen in Fig. 2, are not seen in Fig. 4 and also the redshifts have been decreased. 

If we choose a longer trapezoidal laser pulse (without changing the optical cycles of the rising and falling parts of the laser pulse), we expect that contribution of laser falling part  becomes less as the HHG process occurs mostly before laser falling part especially for the lighter isotopes with faster nuclear motions and more population depletion. To show this, we calculated the HHG spectra for ten-cycle laser pulses (as shown in Fig. 1) with other laser parameters similar to those of Fig. 2. The results are shown in Fig. 5 with their corresponding time profiles depicted in Fig. 6. As can be seen in Fig. 5, the redshifts are small in comparison with Fig. 2 and for H$_2^+$ we can say that no redshift is observed but for D$_2^+$ and X$_2^+$ small redshifts are seen. Fig. 6 shows that for H$_2^+$ and D$_2^+$ most HHG occurs before the pulse falling part. 
The intensity of harmonic peaks decreases for H$_2^+$ and D$_2^+$ at time $t>6.5$ o.c. and at  $t>7.5$ o.c., respectively. 
The population of H$_2^+$ (D$_2^+$) is depleted from 0.8 at $t=6.5$ o.c. ($7.5$ o.c.) to 0.15 (0.45)
at the end of the ten-cycle laser pulse .
Therefore, these  population depletions are responsible for the suppression of the HHG at the last two-cycle laser pulse in Fig. 6. For X$_2^+$, the population depletion is minimum due to slow nuclear motion not allowing the nuclei reach large internuclear distances where enhanced ionization occurs. Therefore the HHG has considerable intensity in the falling part of the laser pulse.
Fig. 6 shows that for X$_2^+$ there are three HHG peaks  having redshift at the falling part of the laser pulse (between 8-10 optical cycles) and there are twelve HHG peaks without redshift before the falling part (between 2-8 optical cycles) which reduce and cover the contribution of the three peaks with redshift into the total HHG spectrum as we do not observe any redshift in the total HHG spectrum of X$_2^+$ in Fig. 5.
In Fig. 7, the ionized outgoing wavepacket (IOW) probability density in the $z$ direction as a function of internuclear distance  for the ten-cycle laser pulse is shown for H$_2^+$, D$_2^+$ and X$_2^+$. This figure shows that the ionization occurs for D$_2^+$ and H$_2^+$ mostly at large internuclear distances where charge-resonance enhanced ionization happens. 
It must be noted for Fig. 7 that we calculated the IOW probability density  from the boundary on the electronic z coordinate by the virtual detector methods [32]. The nuclei in this IOW are mainly under influence of the Coulomb repulsion of each other. Therefore, we expect the internuclear distance of the IOW increases between the ionization time and the time to reach the virtual detector. This increase of internuclear distance is more for a smaller internuclear distance. Therefore, if we want to introduce the IOW probability density in Fig. 7 as the ionization probability, it is relatively deformed and shifted to larger internuclear distances.

The time-dependent population as function of internuclear distance is also shown in Fig. 8 for the H$_2^+$, D$_2^+$ and X$_2^+$ for the ten-cycle laser pulse. The probability of finding the molecule at larger internuclear separations increases in time which is more for the lighter isotopomers. The intensity decrease of harmonic peaks in Fig. 6 for H$_2^+$ ($t>6.5$ o.c.) and D$_2^+$ ($t>7.5$ o.c.) is consistent with finding the molecule at large internuclear distances (Fig. 8) where ionization (Fig. 7) and therefore the population depletion is maximum.

In order to figure out the contributions of different electronic and vibrational states into the redshifts observed in Fig. 2 for the seven-cycle laser pulse, we plotted separately recombination to dominant electronic states, $1s\sigma_g$ and $2p\sigma_u$, and different vibrational states ($\psi_{\nu_i}$) of $1s\sigma_g$ state which are depicted in Figs. 9-11 for X$_2^+$, D$_2^+$ and H$_2^+$, respectively.
 We should note that for Figs. 9-11 we have depicted harmonic orders greater than 19 to focus mainly on the recombination of continuum electron into bound states. Transitions between bound electronic states also lead to generation of harmonic orders up to ionization potential $I_p$ (in this work $I_p=1.1$ corresponding to harmonic order $\sim$ 19), leading to further complexity of harmonic orders below $I_p$ which was not intended in this work.
In Fig. 9(a) for X$_2^+$, the spectra of $S_g+S_u$ and  $S_{tot}$ have nearly the same magnitude and thus we can  conclude that EIT has not significant effect. It is also observed for recombinations to $1s\sigma_g$ and $2p\sigma_u$ states that $S_g$ is dominant over $S_u$ for almost high harmonic orders but at low harmonic orders $S_u$ is slightly higher (Fig. 9(b)). As $S_g$ is dominant for most harmonic orders, so we focus on its components, $S_b$ and $S_d$, to  find out their contribution in the redshifts seen in Fig. 2. As can be seen in Fig. 9(c), $S_d$ is much lower than $S_b$ and consequently dissociative part of $S_g$ has insignificant portion. Comparison of $S_g$ and $S_b+S_d$ in Fig. 9(b) indicates that the VIT is negligible. Therefore vibrational components of $S_b$ should be the source of  redshifts observed for X$_2^+$. To find out how many vibrational states contribute to $S_b$, we compared different $S_b^n$ with $S_b$ and found that $S_b^{10}$ matches well with the $S_b$, i.e. mainly up to 10 vibrational states contribute to the $S_b$. Therefore, lowest vibrational states ($\nu_i\leq$9) mainly contribute to the $S_{tot}$ and the higher-lying vibrational states ($\nu_i>$9) have little contribution. It was expected because of the heavy nuclei of X$_2^+$ which makes nuclear motion slower and thus the nuclear wavepacket can not reach large internuclear distances at these seven- and ten-cycle laser pulses and therefore high vibrational states are not significantly populated. We also already observed that $S_d$ (Fig. 9(c)) was also very weak which is consistent with contribution of the lowest bound vibrational states discussed above. If we compare $S_b^{2}$ and $S_b^{3}$ in Fig. 9(d), the redshifts are larger for the $S_b^{3}$ demonstrating that the redshifts in the HHG spectra originate from the contribution of higher vibrational states of $1s\sigma_g$ state. The patterns of the HHG spectra for $S_b^{4}$ to $S_b^{10}$ is similar to those of $S_b^{3}$ and only the HHG signals have been intensified, with no extra shift compared to what observed for $S_b^{3}$. 
 	
Now we investigate contributions of decomposed components of total HHG spectrum for D$_2^+$ in Fig. 10 as applied for the X$_2^+$. The D$_2^+$ is much lighter than X$_2^+$ and therefore we expect a faster nuclear dynamics and different coupled electron-nuclear dynamics.  By comparing $S_g+S_u$ and  $S_{tot}$ in Fig. 10(a), it can be deduced that the EIT is not important for low harmonic orders but has a pronounced effect on high ones ($>$39), leading to considerable modulations on the signal. The EIT weakened the signal, partially around odd-order harmonics and remarkably around even-order harmonics.
In Fig. 10(b), for low harmonic orders ($<$39), the intensities of $S_g$ and $S_u$ differ considerably which is the reason  why the EIT is insignificant at these harmonic orders. But for harmonics greater than 39, $S_g$ and $S_u$ have almost close signal magnitude and this is while, as mentioned above, the EIT is more noticeable for these harmonic orders. 
 One of the important points observed for $S_g$ and $S_u$ is that the redshifts occur for both components.  The $2p\sigma_u$ state is an anti-bonding (dissociative) state and  this pure dissociative electronic state has also contributed to the observed redshifts.
Similar to X$_2^+$, $S_d$ is much less intense than $S_b$ and thus its contribution to $S_g$ is negligible. Comparing $S_g$ with $S_b+S_d$ in Fig. 10(b) reveals that the VIT is insignificant similar to what observed for $X_2^+$. Fig. 10(d) shows that the number of  vibrational states contributing to the $S_b$ has been increased in comparison with X$_2^+$.
 The bound vibrational stationary states of the electronic ground state of X$_2^+$ are more and more compact than those of D$_2^+$, thus contribution of more vibrational states of the latter into the HHG indicates that the D$_2^+$ has reached larger internuclear distances. That means that vibrational states with higher energies (wider nuclear distribution) are populated for the D$_2^+$ and therefore contribute to the HHG. 
  Fig. 10(d) shows that with increasing the number of $\nu_i$'s in $S_b$ (Eq. 10), the redshift of odd harmonics has become larger and even harmonic orders has gotten more intense due to the contribution of higher vibrational states. Contribution of high vibrational states to the formation of even harmonic orders has been also recently reported for a one-dimensional H$_2^+$ in relatively long laser pulses (10- and 14-cycle pulses) [18]. We mentioned above that the EIT has dominant effect for high harmonic orders (Fig. 10(a)). 
Another point that we can add about the EIT is that its effect, as already mentioned, is seen to be more noticeable around even harmonic orders rather than those close to odd harmonic orders. As contribution to even-generated harmonic orders comes mainly from high-lying vibrational states and these states are closer in energy to the $2p\sigma_u$ state than low vibrational states, the EIT affects considerably the HHG signal around even harmonic orders than those close to odd harmonic orders.
     
Most above justifications for observations in D$_2^+$ can be considered for H$_2^+$ and we just note some specific features of the latter.
The EIT has increased the signal for low harmonic orders but has decreased it for high ones (Fig. 11(a)). The signal $S_u$ is higher than $S_g$ for harmonic orders 39-53 and the redshifts are seen for both $S_g$ and $S_u$ in Fig. 11(b). Due to faster nuclear motion of H$_2^+$ relative to D$_2^+$ and X$_2^+$, the molecular ion can reach large internuclear distances in which charge-resonance enhanced ionization occurs and leads to considerable population in $2p\sigma_u$ state. Unlike X$_2^+$ and D$_2^+$, $S_d$ shows comparable intensity as $S_b$ and even for some high harmonic orders is higher than $S_b$ as shown in Fig. 11(c). Comparing $S_g$ with $S_b+S_d$ in Fig. 11(b) shows that the VIT could change (weaken or enhance) the HHG spectrum not observed for D$_2^+$ and X$_2^+$. In Fig. 11(d), as the number of vibrational states that contribute to $S_b^n$ increases, the peaks close to odd harmonics are slightly being red-shifted and the HHG spectrum near even harmonics get more intense. 

For the case of the ten-cycle laser pulse, all the HHG components, $S_{tot}$, $S_g$, $S_u$, $S_b$, $S_d$ and $S_b^n$'s show only odd harmonic orders with very small redshift compared to those observed for the seven-cycle laser pulse in Fig. 2. The contribution of the above-mentioned decomposed components for each isotopomer is similar (only around odd harmonic orders) to those observed and mentioned in Figs. 9-11. The only difference seen is that the  VIT is insignificant for the ten-cycle laser pulse case, even for the H$_2^+$.

% \begin{table}tio===
% \caption{\label{Table}
%The cutoff harmonic order $N_c$ according to the three-step model for H$_2^+$, with ionization potential $I_p=1.1$ a.u., and pondermotive energy $U_p$ under 800 nm wavelength ($\omega_l=0.057$ a.u.) and $I$=4, 5, 7, 10 $\times 10^{14}$ W$/$cm$^2$ intensities. The effective intensity (of the two central peaks, $I_e$) due to Gaussian envelope is also given for each laser intensity, $I$. The $N_c$ and $U_p$ are calculated for effective intensities.
%} es
% \begin{tabular}{|c|c|c|}
%\hline
% $I (I_e)$ W/cm$^2$& $U_p\, (a.u.)$ & $N_c=(3.17U_p+1.32I_p)/\omega_l$ \\ 
% \hline 
% 4 (3.15)$\times 10^{14}$ & 0.69 & 64 \\ 
% \hline 
% 5 (3.9)$\times 10^{14}$& 0.86 & 73 \\ 
% \hline 
% 7 (5.5)$\times 10^{14}$& 1.21 & 93 \\ 
% \hline 
% 10 (7.8)$\times 10^{14}$ & 1.71 & 121 \\ 
% \hline
% %\label{
%%\label{PIR}
% \end{tabular} 
% \end{table}

\section{Conclusion}
We solved numerically full-dimensional electronic time-dependent Schr\"{o}dinger equation for H$_2^+$ and its isotopomers with and without Born-Oppenheimer approximation under seven- and ten-cycle trapezoidal laser pulses of 800 nm wavelength and $I$=4$\times 10^{14}$ W$/$cm$^2$ intensity. The effects of the falling part of trapezoidal laser pulses and control of these effects on the HHG spectrum were investigated by considering different pulse durations and isotopomers. 

In the case of the fixed-nuclei calculation, only odd harmonic orders are generated but for the case of the freed nuclei (beyond the Born-Oppenheimer approximation), the HHG spectra change depending on the isotopomer. The results show that considerable redshifts appear for a shorter laser pulse (seven cycles) in the presence of nuclear motion, originating mainly from the last two-cycle falling part of the laser pulse. 
Whereas, by considering a longer laser pulse (ten cycles), the redshifts and complexity of the HHG spectra are considerably reduced and controlled because the HHG at the falling part of the laser pulse has a little contribution on the total HHG spectrum. For D$_2^+$ and H$_2^+$ under ten-cycle laser pulses, the HHG is suppressed considerably at the laser falling part due to the population depletion. 

In order to get insight into the underlying components of the complicated patterns observed for the seven-cycle trapezoidal laser pulses, the HHG spectrum was decomposed  into different components representing recombinations to different electronic states $1s\sigma_g$ and $2p\sigma_u$. We also decompose recombination to ground electronic state $1s\sigma_g$ into the bound ($S_b$) and dissociative ($S_d$) components. 
 For the heaviest isotopomer, X$_2^+$, mostly ground electronic state and low vibrational states contribute to the HHG spectrum and high harmonic orders show a redshift coming mainly from the high vibrational states of the $1s\sigma_g$ state. 
 For isotopomers D$_2^+$ and H$_2^+$, recombination to the first excited electronic state $2p\sigma_u$ becomes also significant and the interference term (EIT) between $1s\sigma_g$ and $2p\sigma_u$ states plays an important role and changes the HHG signal, remarkably around even-order harmonics.    
 More vibrational states contribute to the HHG signal for lighter isotopomers indicating that larger internuclear distances have become accessible during the interaction due to the faster nuclear motion. 
  With increasing the number of vibrational states, the HHG signal of odd harmonic orders is being slightly redshifted and that of even-order harmonics gets more intense.  The dissociation part $S_d$ and VIT seen to be noticeable for the H$_2^+$.
  
   Another way that we can suppress the effect of the falling laser pulse on the HHG spectrum is to use enough high laser peak intensities, instead of a longer laser pulse, not considered in this work. We should note that, at very high laser intensities or longer laser rising parts, the blueshift of harmonics at the laser rising part may arise which should be considered. 
 
  \FloatBarrier 
%=====================================Refrences=======
\section{References}
\bibliography{p7}% Produces the bibliography via BibTeX.

\begin{thebibliography}{36}
\bibitem{1} 
T. Brabec and F. Krausz, Rev. Mod. Phys. \textbf{72}, 545 
(2000). 

%
\bibitem{2}
C. Winterfeldt, C. Spielmann, and G. Gerber, Rev. Mod. Phys. 
\textbf{80}, 117 (2008). 
 
\bibitem{3}
 P. B. Corkum, Phys. Rev. Lett. \textbf{71}, 1994 (1993).

\bibitem{4}
M. Lewenstein, P. Balcou, M. Y. Ivanov, A. L$^{^,}$Huillier and P. A. Corkum, Phys. Rev. A \textbf{49}, 2117 (1994).


\bibitem{5}
F. Krausz and M. Y. Ivanov, Rev. Mod. Phys. \textbf{81}, 163 (2009).



\bibitem{6}
J. Itatani, J. Levesque, D. Zeidler, H. Niikura, H. P\'{e}pin, J. C.
Kieffer, P. B. Corkum, and D. M. Villeneuve, Nature (London)
\textbf{432}, 867 (2004).

\bibitem{7}
S. Baker, J. S. Robinson, C. A. Haworth, H. Teng, R. A.
Smith, C. C. Chiril$\breve{a}$, M. Lein, J. W. G. Tisch, and J. P.
Marangos, Science \textbf{312}, 424 (2006).

\bibitem{8}
S. Haessler, J. Caillat, W. Boutu, C. Giovanetti-Teixeira,
T. Ruchon, T. Auguste, Z. Diveki, P. Breger, A. Maquet,
B. Carr\'{e}, R. Ta\"{i}eb, and P. Sali\`{e}res, Nat. Phys. \textbf{6}, 200 (2010).

\bibitem{9}
C. Vozzi, M. Negro, F. Calegari, G. Sansone, M. Nisoli, S. D.
Silvestri, and S. Stagira, Nat. Phys. \textbf{7}, 822 (2011). 

\bibitem{10}
 W. Li, X. Zhou, R. Lock, S. Patchkovskii, A. Stolow, H. C.
Kapteyn, and M. M. Murnane, Science \textbf{322}, 1207 (2008). 

\bibitem{11}
M. Lein, Phys. Rev. Lett. \textbf{94}, 053004 (2005).

\bibitem{12}
X. B. Bian and A. D. Bandrauk, Phys. Rev. Lett. \textbf{113}, 193901 (2014).

\bibitem{13}
S. Baker, J. S. Robinson, M. Lein, C. C. Chiril$\breve{a}$, R. Torres, H. C.
Bandulet, D. Comtois, J. C. Kieffer, D. M. Villeneuve, J. W. G.
Tisch, and J. P. Marangos, Phys. Rev. Lett. \textbf{101}, 053901 (2008).

\bibitem{14}
H. Mizutani, S. Minemoto, Y. Oguchi and
H. Sakai, J. Phys. B: At. Mol. Opt. Phys. \textbf{44},  081002 (2011).

\bibitem{15}
A. D. Bandrauk, S. Chelkowski and H. Lu, J. Phys. B: At. Mol. Opt. Phys. \textbf{42}, 075602 (2009).

\bibitem{16}
L. Feng and T. Chu, J. Chem. Phys. \textbf{136}, 054102 (2012).


\bibitem{17}
X. L. Ge, T. Wang, J. Guo, and X. S. Liu Phys. Rev. A \textbf{89}, 023424 (2014).

\bibitem{18}
F. Morales, P. Rivi\`{e}re, M. Richter, A. Gubaydullin,
M. Ivanov, O. Smirnova and F. Mart\'{i}n, J. Phys. B: At. Mol. Opt. Phys. \textbf{47}, 204015 (2014).

\bibitem{19}
G. Castiglia, P. P. Corso, R. Daniele, E. Fiordilino, B. Frusteri and
F. Morales, Laser Phys. \textbf{23}, 095301 (2013).

\bibitem{20}
Y. C. Han and L. B. Madsen Phys. Rev. A \textbf{87}, 043404 (2013).


\bibitem{21}
H. Ahmadi, A. Maghari, H. Sabzyan, A. R. Niknam and M. Vafaee, Phys. Rev. A \textbf{90}, 043411 (2014).


\bibitem{22}
C. Vozzi, F. Calegari, E. Benedetti, J.-P. Caumes, G. Sansone, S. Stagira, and M. Nisoli, R. Torres, E. Heesel, N. Kajumba, and J. P. Marangos, C. Altucci and R. Velotta, Phys. Rev. Lett. \textbf{95}, 153902 (2005).

\bibitem{23}
T. Kawai, S. Minemote and A. Sakai, Nature \textbf{435}, 470 (2005).

\bibitem{24}
K. B. McFarland, P. J. Farrell, P. H. Bucksbaum and M. G\"{u}hr, Science \textbf{322} 1232 (2008).


\bibitem{25}
S. Fleischer, Y. Zhou, R. W. Field, and K. A. Nelson,
Phys. Rev. Lett. \textbf{107}, 163603 (2011).

\bibitem{26}
D. Shafir, H. Soifer, B. D. Bruner, M. Dagan, Y. Mairesse, S. Patchkovskii, M. Y. Ivanov,
O. Smirnova and N. Dudovich,  Nature \textbf{485}, 343 (2012).

\bibitem{27}
C. T. L. Smeenk and P. B. Corkum,  J. Phys. B: At. Mol. Opt. Phys. \textbf{46}, 201001 (2013).

\bibitem{28}
J. R. Hiskes, Phys. Rev. \textbf{122}, 1207 (1961). 


\bibitem{29}
M. Vafaee, Phys. Rev. A \textbf{78}, 023410 (2008).

\bibitem{30}
A. D. Bandrauk and H. Shen, J. Chem. Phys. \textbf{99}, 1185 (͑1993)͒.

\bibitem{31}
 M. D. Feit, J. A. Fleck, Jr. , and A. Steiger, J. Comput. Phys. \textbf{47}, 412 (1982).
 
\bibitem{32}
M. Vafaee and H. Sabzyan, J. Phys. B \textbf{37}, 4143 (2004).

\bibitem{33}
M. Vafaee, H. Sabzyan, Z. Vafaee, and A. Katanforoush,
e-print arXiv:physics/0509072.


\bibitem{34}
M. Vafaee, H. Sabzyan, Z. Vafaee, and A. Katanforoush,
Phys. Rev. A \textbf{74}, 043416 (2006).

\bibitem{35}
C. Chandre, S. Wiggins and T. Uzer, Phys. D \textbf{181}, 171 (2003).

\bibitem{36}
A. D. Bandrauk, S. Chelkowski and H. Lu, Chem. Phys. \textbf{414}, 73 (2013).


\end{thebibliography}

%\clearpage
\end{document}